# Features of incommensurate phases in multiferroics

**Bahruz Gadjiev**[*,1]

[1] International University for Nature, Society and Man, 19 Universitetskaya str., Dubna, 141980, Russia
**PACS** 05.70.Jk, 42.65.-k, 81.40.Tv

---

[*] Corresponding author: e-mail gadjiev@uni-dubna.ru, Phone: +7 49621 22478, Fax: +7 49621 22464

We present the results of a theoretical analysis of two-dimensionally modulated incommensurate phases in crystals and define the space distribution of magnetization and polarization vector in the incommensurate phase and demonstrate that their averaging in the incommensurate phase period equals zero. We apply symmetry arguments and show that despite the fact that the phase transition in the system is described by reducible representation of the space group high symmetry phase the superspace symmetry group of the incommensurate phase contains all symmetry point groups of the high symmetry phase. We analyze the influence of the fractal distribution of defects on the evolution of the incommensurate phase and show that a phase transition occurs in the incommensurate phase that is accompanied by the inversion loss.

**1 Introduction** In recently discovered multiferroic materials $RMnO_3$ (with the perovskite structure; $R$ — different rare earths, such as $Tb$, $Gd$, $Dy$) and $RMn_2O_5$ ($R = Tb$, $Ho$, $Dy$), as well as in hexaferrites $TbMnO_3$ the ferroelectric order occurs in the magnetically ordered phase [1].

Let's recollect that the magnetoelectric effect permits one to create electric polarization in some systems by applying an external magnetic field. If an external field can do it, it is possible that the same can happen spontaneously due to an internal field or due to certain magnetic ordering [1]. Besides, if we consider the system undergoing phase transition of the second order at temperature $T_c$, applying of a weak external field (magnetic or electrical) can lead to an incommensurate phase with finite temperature width [2]. This supposition allows an extension of the incommensurate phase theory to multiferroics. An intriguing feature of multiferroics is a coexistence the spontaneously induced polarization and magnetic ordering in the incommensurate phases of these compounds. Incommensurate phases can be described in the framework of the Landau phase transitions theory. In the framework of the Landau theory such transitions in multiferroics are described by reducible representation of the space group of the high-symmetry phase. The form of the coupling of electric polarization **P** to magnetization **M** can be found using general symmetry arguments. Namely, ferroelectricity can appear if the spin rotation axis **e** does not coincide with wave vector of a spiral **Q**: the polarization **P** appears only if these two directions are different, and it is proportional to the vector product of **e** and **Q**, **P** ~ [**e**×**Q**] [1].

In doped crystals it is proper inhomogeneity of structure [3]. Presence of inhomogeneities, internal fields and impurities in structure leads to necessity to investigate influence of defects on incommensurable phases.

The paper has the following structure: first, we solve the problem of two-dimensionally modulated incommensurate phases in multiferroics. Then, we perform a symmetry analysis of the incommensurate phases structure. Finally, we analyze the influence of the fractal distribution of defects on incommensurate phases.

**2 Coexistence of magnetic and ferroelectric ordering in 2D modulated incommensurate phases** Let us assume that a crystal with the average space group $G$ in the paramagnetic phase undergoes a magnetic phase transition at critical point $T_c$. Let the order parameter ($\eta_1$, $\eta_2$) (magnetization components) transformed by the active irreducible representation $D$ of the symmetry group of the paramagnetic phase corresponding to point $\mathbf{q_0} = (0,0,0)$ of Brillouin zone. If the compound in the paramagnetic phase has rhombohedral symmetry, the symmetry analysis shows that the following gradient invariants are possible:

$$gP_z\left(\eta_1 \frac{\partial \eta_2}{\partial y} - \eta_2 \frac{\partial \eta_1}{\partial y}\right) \text{ and } \sigma P_y\left(\eta_1 \frac{\partial \eta_2}{\partial z} - \eta_2 \frac{\partial \eta_1}{\partial z}\right),$$

where $P_z$ and $P_y$ are components of the polarization vector. In this case, the symmetry analysis shows that the density of the free energy functional after introducing the variables $\eta_1 = \rho \sin \varphi(y,z)$ and $\eta_2 = \rho \cos \varphi(y,z)$, and using the constant amplitude approximation can be written in the form

$$f = f_0 + f_1, \qquad (1)$$

where

$$f_0 = \frac{\alpha}{2}\rho^2 + \frac{\beta}{4}\rho^4 + \gamma\rho^n \cos n\varphi + \frac{\chi}{2}\left(P_y^2 + P_z^2\right),$$

$$f_1 = -\sigma\rho^2 P_y\left(\frac{\partial \varphi}{\partial z}\right) - gP_z\rho^2\left(\frac{\partial \varphi}{\partial y}\right) + \frac{\tilde{k}}{2}\rho^2\left(\frac{\partial \varphi}{\partial z}\right)^2 + \frac{\tilde{\kappa}}{2}\rho^2\left(\frac{\partial \varphi}{\partial y}\right)^2 + \frac{\nu}{2}\left(\frac{\partial \varphi}{\partial z}\right)\left(\frac{\partial \varphi}{\partial y}\right).$$

We suppose that the parameters of the free energy functional are positive. Here $\alpha = \alpha_0(T - T_c)$, where $T$ is the temperature and $n$ is a degree of anisotropy. We discuss crystal structures that possess heterogeneity as a consequence of the fact that they are formed by various atoms [3]. Due to this, these structures have averaged space group symmetry. Consequently, the free energy density that corresponds to these structures can always contain low order gradient of the order parameter, even if such forms are not allowed by the symmetry.

The free energy functional is presented in the form

$$F = \iint_\Omega f(y,z,\varphi(y,z),\varphi'_y,\varphi'_z)dydz. \qquad (2)$$

From the extreme condition of the free energy functional density for $P_y$ and $P_z$ we obtain $P_y = (\sigma\rho^2/\chi)(\partial\varphi/\partial z)$, $P_z = (g\rho^2/\chi)(\partial\varphi/\partial y)$. After substitution $P_y$ and $P_z$ in (2) we can show that the equation of motion corresponding to functional (2) for the phase of order parameter is represented by the expression

$$k\frac{\partial^2 \varphi(y,z)}{\partial z^2} + \kappa\frac{\partial^2 \varphi(y,z)}{\partial y^2} + \nu\frac{\partial^2 \varphi(y,z)}{\partial y \partial z} + n\gamma\rho^{n-2}\sin n\varphi(y,z) = 0. \qquad (3)$$

Here $k = \tilde{k} - \sigma^2\rho^4/(2\chi)$ and $\kappa = \tilde{\kappa} - g^2\rho^4/(2\chi)$. After introducing the notations $a = k/(n^2\gamma\rho^{n-2})$, $2b = \nu/(n^2\gamma\rho^{n-2})$, $c = \kappa/(n^2\gamma\rho^{n-2})$ and $\psi(y,z) = n\varphi(y,z)$, the equation (3) can be presented as

$$a\frac{\partial^2 \psi(y,z)}{\partial z^2} + 2b\frac{\partial^2 \psi(y,z)}{\partial y \partial z} + c\frac{\partial^2 \psi(y,z)}{\partial y^2} + \sin\psi(y,z) = 0 \qquad (4)$$

Equation (4) is an equation of the hyperbolic type, if $b^2 - ac > 0$ ($\nu^2 - 4k\kappa > 0$), the parabolic type if $b^2 - ac = 0$ ($\nu^2 - 4k\kappa = 0$) and elliptic type if $b^2 - ac < 0$ ($\nu^2 - 4k\kappa < 0$) [4]. After standard transformations [4] equation (6) takes the form

(i) $\dfrac{\partial^2 \psi(\xi,\eta)}{\partial \xi^2} + \dfrac{\partial^2 \psi(\xi,\eta)}{\partial \eta^2} + \sin\psi(\xi,\eta) = 0$, if $\nu^2 - 4k\kappa < 0$;

(ii) $\dfrac{\partial^2 \psi(\xi,\eta)}{\partial \xi^2} - \dfrac{\partial^2 \psi(\xi,\eta)}{\partial \eta^2} - \sin\psi(\xi,\eta) = 0$, if $\nu^2 - 4k\kappa > 0$;  (5)

(iii) $\dfrac{\partial^2 \psi(\zeta)}{\partial \zeta^2} + \sin\psi(\zeta) = 0$, if $\nu^2 - 4k\kappa = 0$.

From the positive determination of the free energy functional parameters it follows that realization (5ii) is physically more probable, and it can be shown that the solution of this equation has the form

$$\psi(\xi,\eta) = 4\arctg\left\{\frac{h'}{k'}sn\left[\left(\frac{16c}{6+4cd}\right)^{1/2}\frac{1}{k'}\xi\right]sn^{-1}\left[\left(\frac{16}{d}\right)\frac{1}{h'}\eta\right]\right\} \qquad (6)$$

In the parabolic case, equation (5iii) is one-dimensional and its solution is represented by expression

$$\psi(\zeta) = 2am(\zeta,\kappa), \ 0 \le \kappa \le 1, \qquad (7)$$

where $am(\zeta,\kappa)$ is the elliptic Jacobi function.

It should be stressed that solutions (6) and (7) define the spatial distribution of the order parameter phase. In this case, the mean value of the order parameter phase is equal to zero. The analysis of expressions for $P_y$ and $P_z$ shows that the mean values of components spontaneous polarization in the incommensurate phase are equal to zero. Therefore in incommensurate phase, the antiferromagnetic ordering in the incommensurate phase coexists with antiferroelectic ordering.

**3 Symmetry analysis** Let us consider the phase transition sequence high-symmetry – incommensurate – commensurate phase described with irreducible representation of the space group symmetry of parent phase [5]. It can be shown that if the symmetry space group in the parent phase contains inversion, the superspace group of the incommensurate phase also contains it. For this we assume that the incommensurate phase is described by the order parameter which is transformed according to irreducible representation $D_0(\boldsymbol{k})$ of the space group symmetry $G$ of the parent phase, where $\boldsymbol{k}$ is an arbitrary representative of the irreducible representation star $\{\boldsymbol{k}^*\}$.

Let $\{Q_{ij}\}$, $i = 1,\ldots, n, j = 1,\ldots, p$ is the order parameter where $n$ is the number of wave vectors in star $\{\boldsymbol{k}^*\}$ of the irreducible representation $D_0(\boldsymbol{k})$ and $p$ is the dimension of its small representation; then dimension of $D_0(\boldsymbol{k})$ is $s = n \times p$. The symmetry group of the incommensurate phase contains those elements of the superspace group that leave the structure invariant. It means that these symmetry elements can be defined from equation

$$Q'_{ij} = Q_{ij}, \quad i = 1,\ldots,n, \ j = 1,\ldots, p \qquad (8)$$

where transformed value $Q'_{ij}$ is defined as

$$Q'_{ij} = e^{i\boldsymbol{q}_i * \boldsymbol{r}} \sum_{km} D_0\left(\boldsymbol{k},\{R|\boldsymbol{t}\}\right)_{ikjm} Q_{km} . \qquad (9)$$

The analysis of equations (8) and (9) shows that if the space group symmetry of high–symmetry phase contains inversion, the superspace group of the incommensurate phase also contains it. This implies that an ordering with the mean value of the order parameter equal to zero can occur in the incommensurate phase.

Let us consider the case when the sequence of phase transitions high-symmetry – incommensurate – commensurate phase is described by the reducible representation $D = D_1 \oplus D_2$ of the symmetry group of the parent phase where $D_1$ and $D_2$ determine transformation properties of two major order parameters. In this case, two situations may occur. If expansion $f$-th symmetrized power of irreducible representation $D_1$ contains the irreducible representation $D_2$, i.e.

$$[D_1(\boldsymbol{k_1})]^f = \ldots + D_2(\boldsymbol{k_2}) + \ldots \qquad (10)$$

the incommensurate phase keeps all point group elements of the space group of high-symmetry phase. If expansion of $[D_1(\boldsymbol{k_1})]^f$ does not contain $D_2(\boldsymbol{k_2})$, the point group symmetry of the incommensurate phase restricted to subgroups of the point group of the high-symmetry phase. It can be stated that in ideal crystals the point group symmetry of the incommensurate phase is strictly fixed in the temperature region of its existence.

In the case we considered it is also possible to verify it in the following way: from the condition of the free energy minimum we define the spontaneous polarization $P$, and after substitution of the obtained result in equation (1) the renormalization of the free energy parameters takes place. Consequently, the superspace group of the incommensurate phase symmetry, defined from the expression for the free energy, turns out to be the same as in the case of the zero spontaneous polarization. Thus, accounting for spontaneous polarization does not change the point group of the incommensurate phase symmetry. As in our case the space group of the paramagnetic phase contains inversion, the superspace group of the incommensurate phase symmetry also contains inversion. It follows from this that in multiferroics in the incommensurate phase where magnetic and ferroelectric orderings co-exist, mean values of magnetization and spontaneous polarization will be equal to zero. The average values of magnetization and spontaneous polarization, different from zero, can be induced by the influence of the structure defects in the incommensurate phase.

**3 The effect of fractal distribution of defects on incommensurate structure** From the structure genesis point of view, the crystals structure formed by different atoms can be regarded as a derivative of the close packages structure [3]. Usually, the symmetry of real derivative structures is lower than that of degenerate close packed crystals constructed from similar spheres. Besides, the derivative composites are prone to various structural imperfections. Let us suppose that there is a space fluctuation of the effective local transition temperature and space fluctuations of crystalline fields in a derivative structure [6]. In this case it is possible to show that the order parameter distribution in the structure is defined with the Tsallis distribution [7, 8]. Using this fact, we can show that the equation of motion of the order parameter has a fractal generalization in the form

$$\frac{\partial^\alpha \psi(x)}{\partial x^\alpha} + \sin\psi(x) = 0 , \qquad (11)$$

where $\partial^\alpha/\partial x^\alpha$ — is the Riesz fractional derivative, $\alpha = 2 - \varepsilon$ and $\alpha \neq 1,3,5,\ldots$. For the analysis of equation (11) we use $\varepsilon$–decomposition [9]. Assume that

$$\psi(x) = \psi_0(x) + \varepsilon\psi_1(x) + \ldots, \qquad (12)$$

Then the fields $\psi_0$ and $\psi_1$ are defined from equation

$$\frac{\partial^2 \psi_0(x)}{\partial x^2} + \sin\psi_0(x) = 0 \qquad (13)$$

and

$$\frac{\partial^2 \psi_1(x)}{\partial x^2} + \psi_1(x)\cos\psi_0(x) + D\psi_0(x) = 0, \qquad (14)$$

respectively. Here

$$D\psi_0(x) = \gamma\psi_0^{(2)}(x) + \frac{\varepsilon}{2}\int_0^\infty \left[\psi_0^{(2)}(x-y) + \psi_0^{(2)}(x+y)\right]dy + \ldots,$$

where $\gamma = 0.577156649\ldots$ is the Euler constant. The solution for equation (13) has the form (7). In the sinusoidal regime of the incommensurate phase at $\kappa \to 0$ the solution of equation (13) has the form

$$\psi_1(x) = 2am(x,\kappa) + n\pi . \qquad (15)$$

Let us solve asymptotically equation (11) in the soliton regime of the incommensurate phase, when the superstructure period becomes large enough at $\kappa \to 1$. In this case

$$\frac{\partial^{2-\varepsilon}\psi(x)}{\partial x^{2-\varepsilon}} \approx \varepsilon\psi(0)x^{-2+\varepsilon} - \frac{\partial\psi(0)}{\partial x}x^{-1+\varepsilon} \qquad (16)$$

and, consequently, equation (11) takes the form

$$\varepsilon\psi(0)x^{-2+\varepsilon} - \varepsilon\frac{\partial\psi(0)}{\partial x}x^{-1+\varepsilon} + \sin\psi(x) = 0. \qquad (17)$$

By using the initial condition $\psi(0) = 0$ and $\partial\psi(0)/\partial x = 1$, we obtain

$$\sin\psi(x) = \varepsilon x^{-1+\varepsilon} \qquad (18)$$

The solution of equation (18) is a monotonously decreasing function which tends to $\psi(x) = \pi k$ at large $x$, where $k$ — is an integer number.

Thus, we may state that in the presence of the fractal distribution of defects in the sinusoidal regime of the incommensurate phase, according to solution (15), the structure periodicity persists. Asymptotic solution of equation (16) shows that the periodicity of this structure is lost with the evolution of the structure in the soliton regime of the incommensurate phase. Thus, if the mean values of the spontaneous polarization and magnetization in the sinusoidal regime of the incommensurate phase are equal to zero, in the soliton regime of the incommensurate phase they are different from zero. Otherwise, it means that in the presence of the fractal distribution of defects during the evolution of the incommensurate phase from the sinusoidal regime to the soliton one the loss inversion of the superspace group of incommensurate phase occurs.

**4 Conclusions** The accomplished analysis allows a conclusion that in multiferroics the mean value of spontaneous polarization and magnetization in the incommensurate phase equals to zero. It is also proved by the symmetry consideration. Namely, if the space group of the paramagnetic phase symmetry contains inversion, the superspace group of the incommensurate phase symmetry also contains it. The fractal distribution of defects leads to the principle change of evolution of the incommensurate phase. For example, in the sinusoidal regime of the incommensurate phase the spatial modulation of the structure still persists. However, in the soliton regime of the incommensurate phase the structure becomes aperiodic and in the vicinity of the transition into the commensurate phase the order parameter phase falls to the commensurate value. It means that the evolution of the incommensurate phase is accompanied by the element loss of the inversion centre of the superspace group of the structure. As a result, spontaneous polarization and magnetization occur in this structure.

### References

[1] D.I. Khomskii, arXiv:cond-mat/0601696v1 (2006).
[2] I.M. Vitebsky, JETP **82**, 357 (1982).
[3] B.R. Gadjiev, Low Temperature Physics **26**, 874 (2000).
[4] A.A. Samarskii and A.N. Tikhonov, Equations of Mathematical Physics (Dover Publ. Inc., New York, 1990).
[5] J.M. Perez-Mato, G. Madariaga and M.J. Tello, Phys. Rev. B **3**, 1234 (1984).
[6] A.P. Levanyuk and A.S. Sigov, Defects and Structural Phase Transition (Gordon and Breach Sci. Publ., New York, 1988).
[7] C. Tsallis, Braz. J. Phys. **29**, 1 (1999).
[8] B.R. Gadjiev, in: Proceedings of the CMD-22, Roma, Italy, 2008 (Europhysics Conference Abstract Volume 32F, ISBN: 2-914771-54-1) CD-ROM (paper TUEp.MAT.6).
[9] V.E. Tarasov and G.M. Zaslavsky, Physica A **283**, 291 (2007).